%% file: main.tex
\documentclass{eptcs}

\providecommand{\event}{HCVS 2014} 
\usepackage{breakurl}             
\usepackage{tikz}
\usepackage{amsthm}
\usepackage{alltt}

\input localdefs
\title{Convex polyhedral abstractions, specialisation and property-based predicate splitting in Horn clause verification\thanks{The research leading to these
  results has received funding from the European Union 7th
  Framework Programme 
  under grant agreement no.
  318337, ENTRA - Whole-Systems Energy Transparency and the Danish Natural Science Research Council grant NUSA: Numerical and Symbolic Abstractions for Software Model Checking. }
}
\author{ Bishoksan Kafle
\institute{Roskilde University\\
Denmark}
\email{kafle@ruc.dk}
\and
John P. Gallagher
\institute{Roskilde University\\
Denmark} 
\institute{IMDEA Software Institute\\ Madrid, Spain}
\email{jpg@ruc.dk}
}
\def\titlerunning{Horn clause verification through constraint specialisation and abstract interpretation}
\def\authorrunning{Bishoksan Kafle and John P. Gallagher}
\begin{document}
\maketitle


\begin{abstract}

We present an approach to constrained Horn clause (CHC) verification combining three techniques: abstract interpretation over a domain of convex polyhedra, specialisation of the constraints in CHCs using abstract interpretation of query-answer transformed clauses, and refinement by splitting predicates.  The purpose of the work is to investigate how analysis and transformation tools developed for constraint logic programs (CLP) can be applied to the Horn clause verification problem. Abstract interpretation over convex polyhedra is capable of deriving sophisticated invariants and when used in conjunction with specialisation for propagating constraints it can frequently solve challenging verification problems.  This is a contribution in itself, but refinement is needed when it fails, and the question of how to refine convex polyhedral analyses has not been studied much. We present a refinement technique based on interpolants derived from a counterexample trace; these are used to drive a property-based specialisation that splits predicates, leading in turn to more precise convex polyhedral analyses. The process of specialisation, analysis and splitting can be repeated, in a manner similar to the CEGAR and iterative specialisation approaches.

\end{abstract}

\input intro
\input convex

\input peconstraint

\input refinement

\input experiments

\input related

\section{Conclusion and Future works}
\label{concl}

We described an iterative procedure for Horn clause verification which interleaves abstract interpretation with specialisation. A specialised set of CHCs is produced first by strengthening the constraints in the given clauses using the results of the abstract interpretation.  Then the procedure terminates if an abstract interpretation of the resulting program is  sufficient to verify the required properties, otherwise, a polyvariant specialisation guided by an abstract counterexample is performed using the inferred constraints as well as interpolated constraints. 

In the future, we would like to find a way of  ensuring progress of refinement, maybe using the powerset polyhedra  domain, and also interface our toolchain with SMT solvers for satisfiability checking and  interpolant generation.

\bibliographystyle{eptcs}
\bibliography{refs}
\end{document}

%% file: localdefs.tex
\newcommand{\true}{\mathsf {true}}
\newcommand{\false}{\mathsf {false}}

\newcommand{\body}{\mathcal{B}}

\newcommand{\C}{{\cal C}}

\newcommand{\lfp}{{\sf lfp}}

\newcommand{\qa}{{\sf qa}}

\newcommand{\vars}{\mathsf{vars}}

\newcommand{\atomconstraints}{\mathsf{atomconstraints}}
\newcommand{\thresholds}{\mathsf{thresholds}}
\newcommand{\query}{\mathsf{q}}
\newcommand{\ans}{\mathsf{a}}
\newcommand{\trace}{\mathsf{tr}}
\newcommand{\constr}{\mathsf{constr}}
\newcommand{\Iproj}{\mathsf{proj}}
\newcommand{\SAT}{\mathsf{SAT}}
\newcommand{\interpolant}{\mathsf{interpolant}}
\newcommand{\unknown}{?}

\def\ll{[\![}
\def\rr{]\!]}

     {\end{tabular}\end{tt}\end{small}}
\def\anno#1{{\ooalign{\hfil\raise.07ex\hbox{\small{\rm #1}}\hfil%
        \crcr\mathhexbox20D}}}

\newtheorem{property}{Property}
\newtheorem{lemma}{Lemma}

%% file: intro.tex
\section{Introduction}

In this paper we explore the use of techniques used in constraint logic program (CLP) analysis and specialisation, for the purpose of CHC verification.  Pure CLP is syntactically and semantically the same as CHC. Unlike CLP, CHCs are not always regarded as executable programs, but rather as specifications or semantic representations of other formalisms. However these are only pragmatic distinctions and the semantic equivalence of CHC and CLP means that techniques developed in one framework are applicable to the other. 

Relevant concepts from CLP include the approximation of the minimal model of a CLP program using abstract interpretation, specialisation of a CLP program with respect to a goal and model-preserving transformation of CLP programs.  Relevant concepts drawn from the CHC verification literature include finding a model of a set of CHCs, property-based abstraction, counterexample generation, and refinement of property-based abstraction using interpolants.

The results shown in the paper are preliminary and much research remains to be done in exploiting the many connections and possibilities for cross-fertilisation between CLP and CHC. The contributions of this paper are:
\begin{itemize}
\item
to demonstrate that abstract interpretation over convex polyhedra is capable of deriving sophisticated invariants, and when used in conjunction with specialisation for propagating constraints it can frequently solve challenging verification problems;
\item
to investigate the problem of refinement of polyhedral abstractions, drawing ideas from 

counterexample-guided refinement.
\end{itemize}

In Section \ref{prelim} we define the basic notation and concepts needed for the verification procedure. Section \ref{convex} reviews the technique of abstract interpretation over convex polyhedra, applied to CLP/CHC, along with the important enhancement of this technique using widening thresholds.  In Section \ref{peconstraint} a procedure for specialisation of CHCs is described, based on query-answer transformations and abstract interpretation.  A simple but surprisingly effective verification tool-chain combining specialisation with abstract interpretation is introduced. Section \ref{refinement} explains how to use a (spurious) counterexample from a failed verification attempt to construct a property-based specialisation using  interpolants. Experimental results and related works are reported in Section \ref{experiments} and Section \ref{relwork} respectively. Finally in Section \ref{concl} we conclude and discuss possible extensions and improvements.

\section{Preliminaries}
\label{prelim}
A CHC is a first order predicate logic formula of the form 
$\forall(\phi \wedge B_1(X_1) \wedge \ldots \wedge B_k(X_k) \rightarrow  H(X))$ ($k \ge 0$),  where $\phi$ is a conjunction of constraints with respect to some background theory, $X_i, X$  are (possibly empty) vectors of distinct variables, $B_1,\ldots,B_k, H$ are predicate symbols, $H(X)$ is the head of the clause and $\phi \wedge B_1(X_1) \wedge \ldots \wedge B_k(X_k)$ is the body.   Sometimes the clause is written $H(X) \leftarrow \phi \wedge B_1(X_1),\ldots, B_k(X_k)$ and in concrete examples it is written in the form \texttt{H :- $\phi$, B$_1$(X$_1$),$\ldots$,B$_k$(X$_k$)}.  In examples,  predicate symbols start with lowercase letters while we use uppercase letters for variables.

In this paper we take the constraint theory to be linear arithmetic with the relation symbols $\le, \ge$, $<, >$ and $=$. There is a distinguished predicate symbol $\false$ which is interpreted as false.  In practice the predicate $\false$ only occurs in the head of clauses; we call clauses whose head is $\false$ \emph{integrity constraints}, following the terminology of deductive databases. Thus the formula $\phi_1 \leftarrow \phi_2 \wedge B_1(X_1),\ldots, B_k(X_k)$ is equivalent to the formula $\false \leftarrow  \neg\phi_1 \wedge \phi_2 \wedge B_1(X_1),\ldots, B_k(X_k)$.  The latter might not be a CHC (e.g. if $\phi_1$ contains $=$) but can be converted to an equivalent set of CHCs by transforming the formula $\neg\phi_1$ and distributing any disjunctions that arise over the rest of the body.  
For example, the formula 
\texttt{X=Y :-  p(X,Y)}
is equivalent to the set of CHCs \texttt{
false :- X>Y, p(X,Y)} and \texttt{false :- X<Y, p(X,Y)}.
Integrity constraints can be seen as safety properties.  For example if a set of CHCs encodes the behaviour of a transition system, the bodies of integrity constraints represent unsafe states.  Thus proving safety consists of showing that the bodies of integrity constraints are false in all models of the CHC clauses.  Figure \ref{exprogram} shows an example set of CHCs (taken from \cite{DBLP:conf/pldi/BeyerHMR07}), modeled over reals containing an integrity constraint, and in this example the problem is to prove that the body of the first clause is unsatisfiable.

\begin{figure}
\begin{alltt}
c1. false:- N>0,I=0,A=0,B=0, l(I,A,B,N).
c2. l(I,A,B,N):-I < N, l_body(A,B,A1,B1), I1 = I+1, l(I1,A1,B1,N).
c3. l(I,A,B,N):- I >=N, A + B > 3 * N.
c4. l(I,A,B,N):- I >=N, A + B < 3 * N.
c5. l_body(A0,B0,A1,B1):- A1 = A0+1, B1 = B0+2.
c6. l_body(A0,B0,A1,B1):- A1 = A0+2, B1 = B0+1.
\end{alltt}
\caption{Example program \emph{t4.pl} \cite{DBLP:conf/pldi/BeyerHMR07}}
\label{exprogram}
\end{figure}

\subsection{The CHC verification problem.} To state this more formally, given a set of CHCs $P$, the CHC verification problem is to check whether there exists a model of $P$. Obviously any model of $P$ assigns false to the bodies of integrity constraints.
 We restate this property in terms of the derivability of the predicate $\false$. Let $P \models F$ mean that $F$ is a logical consequence of $P$, that is, that every interpretation satisfying $P$ also satisfies $F$.

\begin{lemma}\label{model-lemma}
$P$ has a model if and only if $P \not\models \false$.
\end{lemma}

\begin{proof}
Writing $I(F)$ to mean that interpretation $I$ satisfies $F$, we have:
 \[
\begin{array}{lll}
P \not\models \false &\equiv & \mathrm{there~exists~some~interpretation~} I  \mathrm{~such~that~} I(P) \mathrm{~and~}\neg I(\false)\\
&&\mathit{by~definition~of~the~} \models \mathit{~relation}\\
&\equiv&  \mathrm{there~exists~some~interpretation~} I  \mathrm{~such~that~} I(P)\\
&&\mathit{(since~} \neg I(\false) \mathit{~is~true~by~defn.~of~} \false\textit{)}\\
&\equiv& P \mathrm{~has~a~model.}\\
\end{array}
\]
\end{proof}
\noindent
This lemma holds for arbitrary interpretations (only assuming that the predicate $\false$ is interpreted as false), uses only the textbook definitions of  ``interpretation" and ``model" and does not depend on the constraint theory.

The verification problem can be formulated deductively rather than model-theoretically. 
We can exploit proof procedures for constraint logic programming \cite{JaffarM94} to reason about the satisfiability of a set of CHCs.  
Let the relation $P \vdash A$ denote that $A$ is derivable from $P$ using some proof procedure.  If the proof procedure is sound 
then $P \vdash A$ implies $P \models A$, which means that $P \vdash \false$ is a sufficient condition for $P$ to have no model, by Lemma \ref{model-lemma}.  This corresponds to using a sound proof procedure to find or check a counterexample.
On the other hand to show that $P$ does have a model, soundness is not enough since we need to establish $P \not\models \false$. As we will see in Section \ref{techniques} we approach this problem by using \emph{approximations} to reason about the non-provability of $\false$, applying the theory of abstract interpretation \cite{DBLP:conf/popl/CousotC77} to a complete proof procedure for atomic formulas (the ``fixed-point semantics" for constraint logic programs \cite[Section 4]{JaffarM94}).  In effect, we construct by abstract interpretation a proof procedure that is \emph{complete} (but possibly not sound) for proofs of atomic formulas.  With such a procedure,   $P \not\vdash \false$ implies $P \not\models \false$ and thus establishes that $P$ has a model.  
\subsection{Representation of Interpretations}
\label{interpretations}
An interpretation of a set of CHCs is represented as a set of \emph{constrained facts} of the form $A \leftarrow \mathcal{C}$ where $A$ is an atomic formula $p(Z_1,\ldots,Z_n)$ where $Z_1,\ldots,Z_n$ are distinct variables and $\mathcal{C}$ is a constraint over $Z_1,\ldots,Z_n$. If $\mathcal{C}$ is $\true$ we write $A \leftarrow$ or just $A$. The constrained fact $A \leftarrow \mathcal{C}$ is shorthand for the set of variable-free facts $A\theta$ such that $\mathcal{C}\theta$ holds in the constraint theory, and an interpretation $M$ denotes the set of all facts denoted by its elements; $M$ assigns true to exactly those facts.  
$M_1 \subseteq M_2$ if the set of denoted facts of $M_1$ is contained in the set of denoted facts of $M_2$.  

\paragraph{Minimal models.} A model of a set of CHCs is an interpretation that satisfies each clause.  There exists a minimal model with respect to the subset ordering, denoted $M\ll P \rr$ where $P$ is the set of CHCs. $M\ll P\rr$ can be computed as the least fixed point ($\lfp$) of an immediate consequences operator (called $S^{\mathit{D}}_P$ in \cite[Section 4] {JaffarM94}), which is an extension of the standard $T_P$ operator from logic programming, extended to handle the constraint domain $D$.  Furthermore $\lfp(S^{\mathit{D}}_P )$ can be computed as the limit of the ascending sequence of interpretations \emph{$\emptyset, S^{\mathit{D}}_P(\emptyset), S^{\mathit{D}}_P(S^{\mathit{D}}_P(\emptyset)), \ldots$}.  This sequence provides a basis for abstract interpretation of CHC clauses.   

\subsection{Proof Techniques}
\label{techniques}

\paragraph{Proof by over-approximation of the minimal model.} It is a standard theorem of CLP that the minimal model $M\ll P \rr$ is equivalent to the set of atomic consequences of $P$.  That is, $P \models p(v_1,\ldots,v_n)$ if and only if $p(v_1,\ldots,v_n) \in M\ll P \rr$. Therefore, the CHC verification problem for $P$ is equivalent to checking that $\false  \not\in M\ll P \rr$. It is sufficient to find a set of constrained facts $M'$ such that $M\ll P \rr \subseteq M'$, where $\false  \not\in M'$.  This technique is called proof by \emph{over-approximation of the minimal model}. 

\paragraph{Proof by specialisation.} A specialisation of a set of CHCs $P$ with respect to an atom $A$ is the transformation of $P$ to another set of CHCs $P'$ such that $P \models A$ if and only if $P' \models A$.  Specialisation is usually viewed as a program optimisation method, specialising some general-purpose program to a subset of its possible inputs,  thereby removing redundancy and pre-computing statically determined computations.  In our context we use specialisation to focus the verification problem on the formula to be proved.  More specifically, we specialise a set of CHCs with respect to a ``query" to the atom $\false$; thus the specialised CHCs entail $\false$ if and only if the original clauses entailed $\false$.

%% file: convex.tex
\section{Abstract Interpretation over Convex Polyhedra}
\label{convex}

Convex polyhedron analysis (CPA) \cite{DBLP:conf/popl/CousotH78} is a program analysis technique based on abstract interpretation \cite{DBLP:conf/popl/CousotC77}. When applied to a set of CHCs $P$ it constructs an over-approximation $M'$ of the minimal model of $P$, where $M'$ contains at most one constrained fact $p(X) \leftarrow \mathcal{C}$ for each predicate $p$. The constraint $\mathcal{C}$ is a conjunction of linear inequalities, representing a convex polyhedron. 
The first application of convex polyhedron analysis to CLP was by Benoy and King \cite{Benoy-King-LOPSTR96}.
Since the domain of convex polyhedra contains infinite increasing chains, the use of a \emph{widening} operator for convex polyhedra \cite{DBLP:conf/popl/CousotC77,DBLP:conf/popl/CousotH78}  is needed to ensure convergence of the abstract interpretation. Furthermore much research has been done on improving the precision of widening operators.  One techniques is known as widening-upto, or widening with thresholds \cite{Halbwachs-94}.  A threshold is an assertion that is combined with a widening operator to improve its precision. 

Recently, a technique for deriving more effective thresholds was developed \cite{Lakhdar-ChaouchJG11}, which we have adapted and found to be effective in experimental studies. The thresholds are computed by the following method.  Let $S^{D}_P$ be the standard immediate consequence operator for CHCs mentioned in Section \ref{interpretations}. That is, if $I$ is a set of constrained facts, $S^{D}_P(I)$ is the set of constrained facts that can be derived in one step from $I$.  Given a constrained fact $p(\bar Z) \leftarrow \C$, define $\atomconstraints(p(\bar Z) \leftarrow \C)$ to be the set of constrained facts $\{p(\bar Z) \leftarrow C_i \mid \C = C_1 \wedge \ldots \wedge C_k, 1 \le i \le k)\}$.  The function $\atomconstraints$ is extended to interpretations by $\atomconstraints(I) = \bigcup_{p(\bar Z) \leftarrow \C \in I}\{\atomconstraints(p(\bar Z) \leftarrow \C)\}$.

Let $I_{\top}$ be the interpretation consisting of the set of constrained facts $p(\bar{Z}) \leftarrow \true$ for each predicate $p$. We perform three iterations of $S^{D}_P$ starting with $I_{\top}$ (the first three elements of a ``top-down" Kleene sequence) and then extract the atomic constraints. That is, $\thresholds$ is defined as follows.
\[
\thresholds(P) = \atomconstraints(S^{{D}(3)}_P(I_{\top}))
\]
\noindent
A difference from the method in \cite{Lakhdar-ChaouchJG11} is that we use the concrete semantic function $S^{D}_P$ rather than the abstract semantic function when computing thresholds.  The set of threshold constraints represents an attempt to find useful predicate properties and when widening they help to preserve invariants that might otherwise be lost during widening.  See \cite{Lakhdar-ChaouchJG11} for further details.  Threshold constraints that are not invariants are simply discarded during widening.

%% file: peconstraint.tex
\section{Specialisation by constraint propagation}
\label{peconstraint}

We next present a procedure for specialising CHC clauses.  In contrast to classical specialisation techniques based on partial evaluation with respect to a goal, the specialisation does not unfold the clauses at all;  rather, we compute a specialised version of each clause in the program, in which the constraints from the goal are propagated top-down and answers are propagated bottom-up.  The implementation is based on query-answer transformations and abstract interpretation over convex polyhedra.

Let $P$ be a set of CHCs and let $A$ be an atomic formula.   For each clause $H \leftarrow \body$ in $P$ we compute a new clause $H \leftarrow C, \body$ where $C$ is a constraint, yielding a program $P_A$ \emph{specialised} for $A$.  If the addition of $C$ makes the clause body unsatisfiable, it is the same as removing the clause from $P_A$. Clearly $P_A$ may have fewer consequences than $P$ but our procedure guarantees that it preserves the inferability of (constrained instances of) $A$. That is, for every constraint $C$ over the variables of $A$, $P \models \forall(C \rightarrow A)$ if and only if $P_A \models \forall(C \rightarrow A)$. 

The procedure is as follows:  the inputs are a set of CHCs $P$ and an atomic formula $A$.
\begin{enumerate}
\item
Compute a \emph{query-answer transformation} of $P$ with respect to $A$, denoted $P^{\qa}_A$, containing predicates $p^{\query}$ and $p^{\ans}$ for each predicate $p$ in $P$.
\item
Compute an over-approximation of the model of $P^{\qa}_A$, expressed as a set of constrained facts $p^{*}(X) \leftarrow C$, where $*$ is $\query$ or $\ans$. We assume that each predicate $p^{*}$ has exactly one constrained fact in the model (where $C$ is possibly $\mathit{false}$ or a disjunction).
\item
For each clause $p(X) \leftarrow \body$ in $P$, let the model of $p^{\ans}$ be $p^{\ans}(X) \leftarrow C^{\ans}$ (where $X$ is the same tuple of variables in $p(X)$ and $p^{\ans}(X)$).
\item
Replace the clause $p(X) \leftarrow \body$ in $P$ by $p(X) \leftarrow  C^{\ans}, \body$ in $P_A$.
\end{enumerate}
Note that if for some predicate $p$, $C^{\ans}$ is $\false$, then all the clauses for $p$ are removed in $P_A$ as their bodies are unsatisfiable.
We now explain each step in turn.
\subsection{The query-answer transformation}
The query-answer transformation was inspired by -- but is a generalisation of -- the magic-set transformation from deductive databases \cite{Bancilhon-Maier-Sagiv-Ullman}.  Its purpose, both in deductive databases and in subsequent applications in logic program analysis \cite{Debray-Ramakrishnan-94} was to simulate goal-directed (\emph{top-down}) computation or deduction in a goal-independent (\emph{bottom-up}) framework.  Let us define the transformation.

Given a set of CHCs $P$ and an atom $A$, the query-answer program for $P$ wrt. $A$, denoted $P^{\qa}_A$,  consists of the following clauses.  For an atom $A=p(t)$, $A^{\ans}$ and $A^{\query}$ represent the atoms $p^{\ans}(t)$ and $p^{\query}(t)$ respectively.
\begin{itemize}
\item
(Answer clauses). For each clause $H \leftarrow C, B_1,\ldots,B_n$ ($n \ge 0$) in $P$, $P^{\qa}_A$ contains the clause $H^{\ans} \leftarrow  C, H^{\query}, B^{\ans}_1,\ldots,B^{\ans}_n$.
\item
(Query clauses). For each clause $H \leftarrow C, B_1,\ldots,B_i,\ldots,B_n$ ($n \ge 0$) in $P$, $P^{\qa}_A$ contains the following clauses:

$
\begin{array}{l}
B_1^{\query} \leftarrow  C, H^{\query}.\\
\cdots\\
B_i^{\query} \leftarrow  C, H^{\query}, B^{\ans}_1,\ldots,B^{\ans}_{i-1}.\\
\cdots\\
B_n^{\query} \leftarrow C, H^{\query}, B^{\ans}_1,\ldots,B^{\ans}_{n-1}.\\
\end{array}
$
\item
(Goal clause).  $A^{\query} \leftarrow \true$.
\end{itemize}
The program $P^{\qa}_A$ encodes a left-to-right, depth-first computation of the query $\leftarrow A$ for CHC clauses $P$ (that is, the standard CLP computation rule, SLD extended with constraints). This is a complete proof procedure, assuming that all clauses matching a given call are explored in parallel. (Note: the incompleteness of standard Prolog CLP proof procedures arises due to the fact that clauses are tried in a fixed order).  

The relationship of the model of the program $P^{\qa}_A$ to the computation of the goal $\leftarrow A$ in $P$ is expressed by the following property\footnote{
Note that the model of $P^{\qa}_A$ might not correspond exactly to the calls and answers in the SLD-computation, since the CLP computation treats constraints as syntactic entities through decision procedures and the actual constraints could differ. 
}. 
An SLD-derivation in CLP is a sequence $G_0,G_1,\ldots,G_k$ where each $G_i$ is a goal $\leftarrow C,B_1,\ldots,B_m$, where $C$ is a constraint and $B_1,\ldots,B_m$ are atoms. In a left-to-right computation, $G_{i+1}$ is obtained by resolving $B_1$ with a program clause. 

\begin{property}[Correctness of query-answer transformation]\label{prop1}
Let $P$ be a set of CHCs and $A$ be an atom.  Let $P^{\qa}_A$ be the query-answer program for $P$ wrt. $A$.  Then 
\begin{itemize}
\item [(i)] if there is an SLD-derivation $G_0, \ldots,G_i$ where $G_0 = \leftarrow A$ and  $G_i = \leftarrow C,B_1,\ldots,B_m$, then $P^{\qa}_A \models \forall (C\vert_{\vars(B_1)} \rightarrow B_1^{\query})$;
\item [(ii)] if there is an SLD-derivation $G_0, \ldots,G_i$ where $G_0 = \leftarrow A$, containing a sub-derivation $G_{j_1},\ldots,G_{j_k}$, where $G_{j_i}\leftarrow C',B_1,B'$ and $G_{j_k} = \leftarrow C,B'$, then $P^{\qa}_A \models \forall (C\vert_{\vars(B_1)} \rightarrow B_1^{\ans})$. (This means that the atom $B_1$ in $G_{j_i}$ was successfully answered, with answer constraint $C\vert_{\vars(B_1)}$).
\item [(iii)] As a special case of (ii), if there is a successful derivation of the goal $\leftarrow A$ with answer constraint $C$ then  $P^{\qa}_A \models \forall (C \rightarrow A^{\ans})$.
\end{itemize}
\end{property}
\noindent

\subsection{Over-approximation of the model of the query-answer program $P^{\qa}_{\false}$}

The query-answer transformation of $P$ with respect to $\false$ is computed. It follows from Property \ref{prop1}(iii) that if $\false$ is derivable from $P$ then $P^{\qa}_{\false} \models \false^{\ans}$.  Convex polyhedral analysis of $P^{\qa}_{\false}$ yields an overapproximation of $M\ll P^{\qa}_{\false} \rr$, say $M'$, containing constrained facts for the query and answer predicates. These represent the calls and answers generated during all derivations starting from the goal $\false$. 

\subsection{Strengthening the constraints in $P$}

 We use the information in $M'$ to specialise the original clauses in $P$. Suppose $M'$ contains constrained facts $p^{\query}(X) \leftarrow  C^{\query}$ and $p^{\ans}(X) \leftarrow  C^{\ans}$. If there is no constrained fact $p^{*}(X) \leftarrow  C^{*}$ for some $p^{*}$ then we consider $M'$ to contain $p^{*}(X) \leftarrow  \mathit{false}$. The clauses in $P$ with head predicate $p$ can be \emph{strengthened} using the constraints $C^{\query}$ and $C^{\ans}$.  Namely, for every clause $p(X) \leftarrow \body$ in $P$ (assuming that the constrained facts are renamed to have the same variables $X$) the conjunction $C^{\query} \wedge C^{\ans}$ are added to the body $\body$. The addition of $C^{\query}$ corresponds to propagating constraints ``top-down" (via the calls) while the addition of $C^{\ans}$ represented propagation ``bottom-up" (via the answers). 
Furthermore, note that $C^{\ans} \rightarrow C^{\query}$ since the answers for $p$ are always stronger than the calls to $p$.  Thus it suffices to add the constraint $C^{\ans}$ to $\body$.

Specialisation by strengthening the constraints preserves the answers of the goal with respect to which the query-answer transformation was performed. In particular, in our application we have the following property.
\begin{property}\label{prop2}
If $P$ is a set of CHCs and $P_{\false}$ is the set obtained by strengthening the clause constraints as just described, then
$P \models \false$ if and only if $P_{\false} \models \false$.
\end{property}
The result of strengthening the constraints in Figure \ref{exprogram}, using the query-answer program with respect to the goal \texttt{false}, is shown in Figure \ref{exprogram-spec}.  Note that the constraint in clause \texttt{c4} is strengthened to \texttt{false}.

\begin{figure}
\begin{alltt}
c1. false:- N>0,I=0,A=0,B=0, l(I,A,B,N).
c2. l(A,B,C,D) :- 2*A+ -1*B>=0,-1*A+1*D>0,-1*A+1*B>=0,3*A+ -1*B+ -1*C=0,
                  1*A+ -1*E= -1,l_body(B,C,F,G),l(E,F,G,D).
c3. l(A,B,C,D):- 3*A+ -3*D>0,1*D>0,2*A+ -1*B>=0,-3*A+3*D> -3,
                -1*A+1*B>=0,3*A+ -1*B+ -1*C=0.
c4. l(A,B,C,D):- false.
c5. l_body(A,B,C,D) :- -1*A+2*B>=0, 2*A+ -1*B>=0,
                       1*A+ -1*C= -1,1*B+ -1*D= -2.
c6. l_body(A,B,C,D) :- -1*A+2*B>=0,2*A+ -1*B>=0,1*A+ -1*C= -2,1*B+ -1*D= -1.
\end{alltt}
\caption{Example program \emph{t4.pl} \cite{DBLP:conf/pldi/BeyerHMR07} with strengthened constraints}
\label{exprogram-spec}
\end{figure}

\subsection{Analysis of the model of the specialised clauses}
It may be that the clauses $P_{\false}$ do not contain a clause with head $\false$.  In this case safety is proven, since clearly $P_{\false} \not\models \false$. If this check fails, the convex polyhedral analysis is now run on the clauses $P_{\false}$. As the experiments later show, safety is often provable by checking the resulting model;  if no constrained fact for $\false$ is present, then 
$P_{\false} \not\models \false$.  If safety is not proven, there are two possibilities: the approximate model is not precise enough, but $P$ has a model, or there is a proof of $\false$.  To distinguish these we proceed to try to refine the clauses by splitting predicates.

%% file: refinement.tex
\section{ Safety Check and Program Refinement}
\label{refinement}

This section outlines a procedure for safety check, counterexample analysis and refinement. Refinement is considered when a proof of safety or an existence of a real counterexample (that is, a proof of $\false$ cannot be established. 
\paragraph{Safety check and counterexample analysis}

The absence of a constrained fact for predicate $\false$ in the over-approximation proves that the given set of CHCs is safe. If safety can not be shown, our implementation of the convex polyhedron analysis produces a derivation tree for $\false$ as a trace term which we define formally below. For our program in Figure \ref{exprogram}, the set of constrained facts representing the approximate model is shown below.
\begin{alltt}
f1. l_body(A,B,C,D) :- 1*B+ -1*D>= -2,-1*B+1*D>=1,-1*A+2*B>=0, 2*A+ -1*B>=0,
			\  \  \ \ \ \	 \ \  \  \ \ \ \	 \   \ \ 1*A+1*B+ -1*C+ -1*D= -3.
f2. false :- true.
f3. l(A,B,C,D) :- 1*D>0,2*A+ -1*B>=0,-1*A+1*B>=0,-3*A+3*D> -3, 
			\  \  \ \ \ \	 \ \  \  \ \ 	   3*A+ -1*B+ -1*C=0.
\end{alltt}
Since there is a constrained fact for $\false$, the shortest derivation for it is found, using clause \texttt{c1} followed by clause \texttt{c3}. This will be represented as a \emph{trace term} \texttt{c1(c3)}, which is formally defined below. The idea of trace terms to capture the shape of derivations was introduced by Gallagher and Lafave \cite{Gallagher-Lafave-Dagstuhl}.

\paragraph{AND-trees and trace terms.} Each CHC is associated with an identifier, as shown in Figure \ref{exprogram}. These identifiers are treated as constructors whose arity is the number of non-constraint atoms in the clause body. 
The following definitions of derivations and trace terms is adapted from 
\cite{Gallagher-Lafave-Dagstuhl}.

An \emph{AND-tree} is a tree each of whose nodes is labelled by an
atom and a clause, such that
\begin{enumerate}
\item
each non-leaf node is labelled by a clause
$A \leftarrow C, A_1,\ldots,A_k$ and an atom $A$, and has children labelled by $A_1,\ldots,A_k$,
\item
each leaf node is labelled by a clause $A \leftarrow C$ and
an atom $A$.
\end{enumerate}
We assume that the variables in node labels are renamed appropriately, details are not given here. Any finite derivation corresponds to an AND-tree, and each AND-tree $T$ can be associated with a trace term $\trace(T)$ defined as:
\begin{enumerate}
\item
$c_j$, if $T$ is a single leaf node labelled by the clause of form $A \leftarrow C$ with identifier
 $c_j$; or
\item
$c_i(\trace(T_1),\ldots,\trace(T_{n}))$, if $T$ is labelled by the
clause with identifier by $c_i$, and has subtrees
$T_1,\ldots,T_{n}$.
\end{enumerate}
A trace-term uniquely defines an AND-tree (up to renaming of variables).  The set of constraints of an AND-tree, represented as $\constr(T)$ is 
\begin{enumerate}
\item
$C$, if $T$ is a single leaf node labelled by the clause of form $A \leftarrow C$; or
\item
$C \cup  \bigcup_{i=1..n}(\constr(T_i))$ if $T$ is labelled by the
clause $A \leftarrow C, A_1,\ldots,A_k$ and has  subtrees
$T_1,\ldots,T_{n}$.
\end{enumerate}
We say that an AND-tree $T$ is satisfiable if $\SAT(\constr(T))$. Let $T$ be an AND-tree whose root is labelled by atom $A$.  Define $\Iproj(T)$ to be $\constr(T) \vert_{\vars(A)}$. 

\paragraph{Interpolants.}  Given two sets of constraints $C_1, C_2 $ such that $C_1 \cup C_2 $ is unsatisfiable, a (Craig) interpolant  is a constraint $I$ with
(1) $C_1 \subseteq I$, (2) $I \cup C_2 $ is unsatisfiable  and (3) $I$ contains only variables common to $C_1$ and $C_2$.  We implemented the algorithm from \cite{DBLP:journals/jsc/RybalchenkoS10} for interpolants for linear constraints.

%
%
%
Given an AND-tree $T$ where $\neg\SAT(\constr(T))$,  we can construct an interpolant for each non-root node of $T$, also known as tree interpolants.  Let $T'$ be a sub-tree of $T$, whose root is labelled with $A'$.  Then the interpolant $I$ associated with $A'$ is defined as above where $C_1 = \constr(T')$ and $C_2 = \constr(T) \setminus C_1$, and the interpolants of subtree of $T'$ together with the constraints at the root of $T'$ implies $I$.  Note that by construction of the AND-tree, the only variables in common between $C_1$ and $C_2$ (and hence in $I$) are the variables in $A'$, the label of $T'$. More details on tree interpolation can be found in \cite{DBLP:conf/lpar/BlancGKK13}.  

The set $\interpolant(T)$ is the set of constrained facts $A \leftarrow I$, for all non-root nodes of $T$ labelled by atom $A$ with interpolant $I$ as defined above.

\paragraph{Counterexample checking.}
Given a trace term, let $T$ be the corresponding AND-tree.  We report that the CHCs have no model if $\SAT(\constr(T))$, and  our procedure terminates.  For our example it can be verified that $\SAT(\constr(\mathtt{c1(c3)}))$  does not hold, so the trace \texttt{c1(c3)} is a false alarm. We now use the interpolants to split predicates and try to get a more precise approximation of the model.

From the trace term \texttt{c1(c3)} in the running example  we derive $\interpolant(\texttt{c1(c3)})= \{I\}$ where $I = \mathtt{l(A,B,C,D) } \leftarrow \mathtt{A+ -3*B+C+D=<0}$.

We then split the constrained facts in the approximation of the model, using the corresponding interpolants and their negations. In the example we split constrained fact \texttt{f3} by strengthening its constraint with $I$ and $\neg I$ respectively. Fioravanti \emph{et al.} use a related technique for splitting clauses \cite{Fioravanti02specializationwith}.
Strengthening first with $I$ we get
\begin{alltt}
l(A,B,C,D):- D>0,2*A+ -1*B>=0,-1*A+1*B>=0,-3*A+3*D> -3,
             3*A+ -1*B+ -1*C=0,A+ -3*B+C+D=<0
\end{alltt}
which after simplification becomes 
\begin{alltt}
l(A,B,C,D) :- -4*A+4*B+ -1*D>=0,1*D>0,-3*A+3*D> -3,2*A+ -1*B>=0,
              3*A+ -1*B+ -1*C=0.
\end{alltt}
We follow the same step with $\neg I$ and obtain the following set of constrained facts.
\begin{alltt}
l(A,B,C,D) :- -4*A+4*B+ -1*D>=0,1*D>0,-3*A+3*D> -3,2*A+ -1*B>=0,
              3*A+ -1*B+ -1*C=0.
l(A,B,C,D) :- 4*A+ -4*B+1*D>0,-1*A+1*B>=0,-3*A+3*D> -3,2*A+ -1*B>=0,
              3*A+ -1*B+ -1*C=0.
\end{alltt}
These together with \texttt{f1} and \texttt{f2}  give us a new  set of constrained facts, which forms the input to the refinement phase of our procedure.
\paragraph {Refinement by Predicate Splitting.} 
Refinement consists of obtaining a specialised set of CHCs from a given set of constrained facts and  input set of CHCs. We do this by using polyvariant specialisation (PS) based on the method of multiple specialisation \cite{Winsborough89} with a property-based abstract domain based on the given set of constrained facts. PS is a program specialisation which introduces several new predicates corresponding to specialised versions of the same predicate. Polyvariant specialisation brings the expressive power of disjunctive predicates into the analysis \cite{DBLP:journals/fuin/FioravantiPPS13}. Space does not permit a more detailed description. For our running example we obtain a split of the predicate \texttt{l} into \texttt{l\_1}  and \texttt{l\_3} , and the specialised program is as follows.

\begin{alltt}
 false :-        1*A>0,1*B=0,1*C=0,1*D=0,l_3(B,C,D,A).
 l_3(A,B,C,D) :- 2*A+ -1*B>=0,-1*A+1*B>=0,-1*A+1*D>0,4*A+ -4*B+1*D>0,
               3*A+ -1*B+ -1*C=0, A+ -1*E= -1,l_body_2(B,C,F,G),l_1(E,F,G,D).
 l_3(A,B,C,D) :- 4*A+ -4*B+1*D>0,3*A+ -3*D>0,-1*A+1*B>=0,-3*A+3*D> -3,
                 3*A+ -1*B+ -1*C=0.
 l_1(A,B,C,D) :- 2*A+ -1*B>=0,-1*A+1*B>=0,-1*A+1*D>0,3*A+ -1*B+ -1*C=0,
                 1*A+ -1*E= -1,l_body_2(B,C,F,G),l_1(E,F,G,D).
 l_1(A,B,C,D) :- 3*A+ -3*D>0,2*A+ -1*B>=0,1*D>0,-1*A+1*B>=0,-3*A+3*D> -3,
                 3*A+ -1*B+ -1*C=0.
 l_body_2(A,B,C,D) :- 2*A+ -1*B>=0,-1*A+2*B>=0,1*A+ -1*C= -1,1*B+ -1*D= -2.
 l_body_2(A,B,C,D) :- 2*A+ -1*B>=0,-1*A+2*B>=0,1*A+ -1*C= -2,1*B+ -1*D= -1.
\end{alltt}

The next iteration continues with this specialised program.  The intention of splitting and PS  is to guarantee progress of refinement, that is, a counterexample once eliminated never occurs again. Our procedure does not guarantee progress, that is, the same spurious counterexamples might appear in subsequent iterations, but in practice we find the polyvariant specialisation usually eliminates the given counterexample. The large number of constants in the above examples are derived during invariants computation.
In the next iteration, our example terminates with a real counter example, thus proving our example program unsafe (over the real numbers).

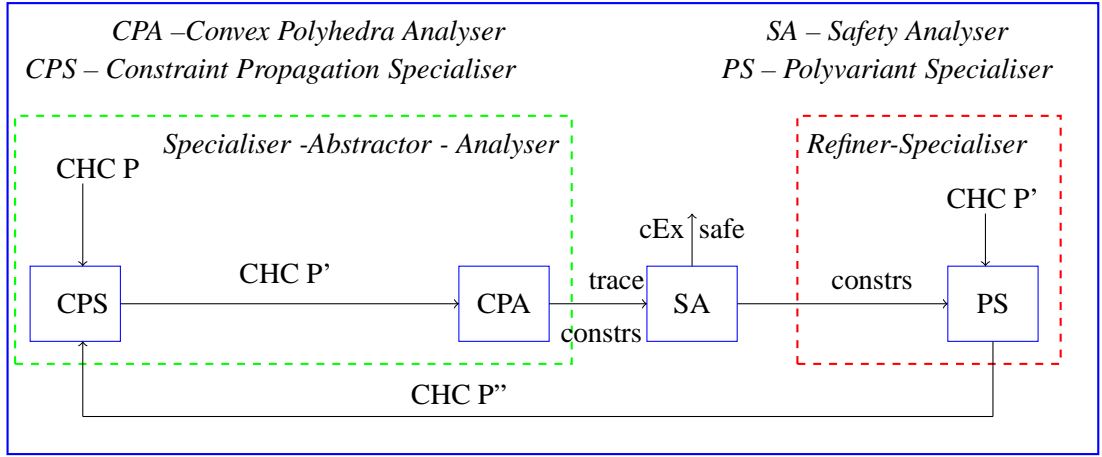
\begin{figure}[h!]
\begin{center}
\begin{tikzpicture}
\draw[thick,  color=blue] (0.5,0) rectangle (15,6); 
\begin{scope}
 \node at (4,5.1) {\it CPS --  Constraint Propagation Specialiser};
\node at (4.5,5.6) {\it CPA --Convex Polyhedra Analyser};
\node at (12.2,5.6) {\it SA -- Safety Analyser};
\node at (12.2,5.1) {\it PS -- Polyvariant Specialiser};

 \node at (1.7,3.8) {CHC  P};
 \draw[->] (1.5,3.6) -- (1.5,2.5);
 
\draw[thick,dashed,green] (.6,1.2) rectangle (8.0,4.5);  
 \node at (5.2,4.1) {\it   Specialiser -Abstractor - Analyser };

\draw[thick,dashed,red] (11,1.2) rectangle (14.5,4.5);  
 \node at (12.6,4.1) {\it   Refiner-Specialiser };

\draw[blue] (.8,1.5) rectangle (2.0,2.5); 
\node at (1.5,2) {CPS}; 

 \draw[blue] (6.5,1.5) rectangle (7.7,2.5); 
\node at (7.1,2) {CPA}; 
\node at (4.2,2.4) {CHC P'}; 
 \draw[->] (2,2) -- (6.5,2);

    \node at (10,3.0) {safe };
   \node at (9.2,3.0) {cEx };
\draw[->] (9.6,2.5) -- (9.6,3.2);


 \draw[->] (7.7,2) -- (9,2);
 \node at (8.6,2.3) {trace};
\node at (8.4,1.6) {constrs};

\node at (12,2.3) {constrs};
 \draw[->] (10.2,2) -- (13,2);

 \node at (6.5,0.8) {CHC P'' };
 \draw[-] (13.6,1.5) -- (13.6,0.5);
  \draw[-] (13.6,0.5) -- (1.5,0.5);
   \draw[->] (1.5,0.5) -- (1.5,1.5);
   
  \draw[->] (13.5,3.2) -- (13.5,2.5);
 \node at (13.6,3.4) {CHC  P'};

\draw[blue] (9,1.5) rectangle (10.2,2.5); 
\node at (9.6,2) {SA }; 

\draw[blue] (13,1.5) rectangle (14.2,2.5); 
\node at (13.6,2) {PS}; 
\end{scope}
\end{tikzpicture}
\end{center}
\caption{\it CHC verification toolchain.}
\label{fig:toolchain}
\end{figure}

\paragraph{Toolchain.}
Our verification procedure is summarised  in Figure \ref{fig:toolchain}, which is divided into three parts, an \emph{abstractor}  (inside green dotted box), followed by a \emph{safety check} and \emph{counterexample analyser} and \emph{refiner} (inside red  box).  It should be noted that the tools inside the green and red boxes produce new set of CHCs by specialisation.

The effects of CPA and PS in our procedure complement each other and the CPA model gets more accurate during refinement which allows generation of better specialised programs. In essence, it marries the effectiveness of CPA with PS.

%% file: experiments.tex
\section{  Experiments}
\label{experiments}

Table \ref{tbl:experiments} presents the results of applying our toolchain depicted in Figure \ref{fig:toolchain}  to a number of benchmark programs  taken  from the repository of Horn clause benchmarks in SMT-LIB2\footnote{https://svn.sosy-lab.org/software/sv-benchmarks/trunk/clauses/} and other sources including \cite{DBLP:journals/tplp/GangeNSSS13,DBLP:conf/rv/JaffarNS11,DBLP:conf/cav/GuptaR09,DBLP:conf/tacas/Beyer13,DBLP:conf/tacas/AngelisFPP14}. The experiments were carried out using a computer, Intel(R)  X5355 having 4 processors (each  @ 2.66GHz) and total memory  of 6 GB. Debian 5 (64 bit)  is the Operating System running in it and we set  2 minutes of timeout for each experiment. Our tool-chain is implemented in 32-bit Ciao Prolog \cite{ciao-reference-manual-tr}\footnote{http://ciao-lang.org/} and the Parma Polyhedra Library \cite{BagnaraHZ08SCP}\footnote{http://bugseng.com/products/ppl/} for this purpose. 

In Table \ref{tbl:experiments}, columns Program, ``n"  , Result and time (sec) respectively represent the benchmark program, the number of refinement iterations necessary to verify a given property, the results of verification and the time (in seconds) to verify them. Value 0 in column ``n" means that no refinement is necessary, whereas value greater than $0$ indicates the actual number of iterations necessary and value ``-" means that these programs are beyond the reach of our current tool within the given time limit. Problems marked with (*) were not  handled by our tool-chain since their solution generates numbers which do not fit in 32 bits, the limit of our Ciao Prolog implementation. Problems such as systemc-token-ring.01-safeil.c contain complicated loop structure with large strongly connected components in the predicate dependency graph and our convex polyhedron analysis tool is unable to derive the required invariant.

\begin{table}[h!]
\begin{minipage}{\textwidth}
    \begin{tabular}{|l|l|l|l|l|l|l|l|}
    \hline
    Program                  & n & Result & time & Program                        & n & Result  & time \\ 
                       &  &  &  (secs) &                         &  &   & (secs) \\ \hline
    MAP-disj.c.map.pl        & 0 & safe   & 1.0     & jaffex1c.pl                    & 0 & safe    & 0.01    \\ \hline
    MAP-forward.c.map.pl     & 0 & safe   & 1.0     & jaffex1a.pl                    & 0 & safe    & 0.01    \\ \hline
    t1.pl                    & 0 & safe   & 0.01    & qrdcmp.smt2                    & 0 & safe    & 118.0   \\ \hline
    t1-a.pl                  & 0 & safe   & 0.01    & choldc.smt2                    & 0 & safe    & 19.0    \\ \hline
    t2.pl                    & 0 & safe   & 0.01    & lop.smt2                       & 0 & safe    & 39.0    \\ \hline
    t3.pl                    & 0 & safe   & 1.0     & pzextr.smt2                    & 0 & safe    & 40.0    \\ \hline
    t4.pl                    & 1 & unsafe & 1.0     & qrsolv.smt2                    & 0 & safe    & 18.0    \\ \hline
    t5.pl                    & 0 & safe   & 0.01    & tridag.smt2                    & 0 & safe    & 13.0    \\ \hline
    MAP-disj.c-scaled.pl & 0 & safe   & 1.0     & systemc-pc-sfifo\_1  & 0 & unsafe  & 12.0    \\ \hline
    INVGEN-id-build          & 0 & safe   & 1.0     & loops-terminator               & 0 & unsafe  & 0.01    \\ \hline
    INVGEN-nested5           & 0 & safe   & 1.0     & loops-for-bounded              & 3 & unsafe  & 5.0     \\ \hline
    INVGEN-nested6           & 0 & safe   & 117.0   & TRACER-testabs15               & 0 & safe    & 1.0     \\ \hline
    INVGEN-nested8           & 0 & safe   & 1.0     & INVGEN-apache-esc-abs  & 0 & safe    & 2.0     \\ \hline
    INVGEN-svdsomeloop     & 0 & safe   & 3.0     & DAGGER-barbr.map.c             & 0 & safe    & 119.0   \\ \hline
    INVGEN-svd1              & 2 & safe   & 13.0    & systemc-token-ring.01-safeil.c & - & \unknown & -       \\ \hline
    INVGEN-svd4              & 0 & safe   & 5.0     & sshs3-srvr1a-safeil.c(*)  & - & \unknown & -       \\ \hline
    loops-count-up-down      & 0 & unsafe & 1.0     & sshs3-srvr1b-safeil.c   & - & \unknown & -       \\ \hline
    loops-sum04              & 8 & unsafe & 2.0     & amebsa.smt2                    & - & \unknown & -       \\ \hline
    dfpp12.pl                & 0 & safe   & 0.01    & bandec.smt2(*)                   & - & \unknown & -       \\ \hline
    TRACER-testloop27        & 1 & unsafe & 1.0     & TRACER-testloop28              & - & \unknown & -       \\ \hline
    TRACER-testloop8         & 0 & unsafe & 0.01    & crank.smt2                     & - & \unknown & -       \\ \hline
    jaffex1b.pl              & 0 & safe   & 0.01    & pldi12.pl                      & - & \unknown & -       \\ \hline
    jaffex1d.pl              & 0 & safe   & 0.01    & loops-sum01                    & - & \unknown & -       \\ \hline
    \end{tabular}
   \end{minipage}
\caption {Experimental results on CHC benchmark problems \label{tbl:experiments}}
\end{table}

The results of our procedure in a larger set of benchmarks obtained from previous sources are summarised in Table \ref{exp:postres}.  Though our tool-chain is not optimized at all, the overall  result shows that it compares favourably with other advanced verification tools like HSF \cite{DBLP:conf/tacas/GrebenshchikovGLPR12}, VeriMAP \cite{DBLP:conf/tacas/AngelisFPP14}, TRACER \cite{DBLP:conf/cav/JaffarMNS12} etc. in both time and the number of problems solved, and thus showing the effectiveness of our approach.
\begin{table}[h!]
\centering
    \begin{tabular}{|c|c|c|}
    \hline
    ~                     &  without refinemet        & with refinement      \\ \hline
    solved  (safe/unsafe) & 160  (142/18) & 181  (158/23) \\ \hline
    unknown/  timeout     & 49/7          & -/35          \\ \hline
    total  time           & 1293          & 3410          \\ \hline
    average  time (secs)         & 5.98          & 18.73         \\ \hline
    \end{tabular}
\caption {Experimental results on 216 CHC verification problems, where ``-" means not relevant. \label{exp:postres}}
\end{table}

%% file: related.tex
\section{Related Work}
\label{relwork}
Verification of CLP programs using abstract interpretation and specialisation has been studied for some time.  The use of an over-approximation of the semantics of a program can be used to establish safety properties -- if a state or property does not appear in an over-approximation, it certainly does not appear in the actual program behaviour.  A general framework for logic program verification through abstraction was described by Levi \cite{Levi00}.

The use of program transformation to verify properties of logic programs was pioneered by Pettorossi and Proietti \cite{PettorossiP00} and Leuschel \cite{LeuschelM99}. Transformations that preserve the minimal model (or other suitable models) of logic programs are applied systematically to make properties explicit.  For example, if a program can be transformed to one containing a clause $A \leftarrow true$ then $A$ is a consequence of the program. 

Recent work by De Angelis \emph{et al.} \cite{DBLP:journals/scp/AngelisFPP14,DBLP:conf/tacas/AngelisFPP14} applies a specialisation approach to the Horn clause verification problem as discussed here, namely, with integrity constraints expressing the properties to be proved.  Both our approach and theirs repeatedly apply specialisations preserving the property to be proved. However the difference is that their specialisation techniques are based on unfold-fold transformations, with a sophisticated control procedure  controlling unfolding and generalisation. Our specialisations are restricted to strengthening of constraints or polyvariant splitting based on local conditions.  Their test for success or failure is a simple syntactic check, whereas ours is based on an abstract interpretation to derive an over-approximation.  

Counterexample guided abstraction refinement (CEGAR)  \cite{DBLP:journals/jacm/ClarkeGJLV03} has been successfully used in  verification  to automatically refine (predicate) abstractions  to reduce false alarms but not much has been explored in refining abstractions in the convex polyhedral domain. 
See \cite{DBLP:conf/sas/BjornerMR13, DBLP:conf/aplas/GuptaPR11} for more details about the use of interpolation in refinement.
A number of tools implementing predicate abstraction and refinement are available, such as HSF \cite{DBLP:conf/tacas/GrebenshchikovGLPR12} and BLAST \cite{DBLP:journals/cacm/BallLR11}. 
TRACER \cite{DBLP:journals/tplp/GangeNSSS13} is a verification tool based on CLP that uses symbolic execution.

 Informally one can say that approaches differ in where the ``hard work"  is performed. In the  work of De Angelis \emph{et al.} the specialisation procedure is the core, whereas in the CEGAR approaches the refinement step is crucial, and interpolation plays a central role.  In our approach, by contrast, most of the hard work is done by the abstract interpretation, which finds useful invariants as well as propagating constraints globally.  The main problem is to find effective ways of refining polyhedral abstractions.  Finding the most effective balance between specialisation, abstraction and refinement techniques is a matter of ongoing research.